\documentstyle[tighten,prl,aps]{revtex}

\begin{document}
\draft
\title{Marginal Fermi Liquid with a Two-Dimensional Patched Fermi Surface}
\author{A.Ferraz$^{*}$,T.Saikawa$^{*}$and Z.Y.Weng$^{\dagger }$}
\address{$*$Laboratorio de Supercondutividade, \\
Centro Internacional de Fisica da Materia Condensada,Universidade de\\
Brasilia,\\
70919-970 Brasilia-DF,Brazil\\
$\dagger $Texas Center for Superconductivity and Department of\\
Physics,University of Houston,Houston,Texas 77204-5506 }
\date{\today}
\maketitle

\begin{abstract}
We consider a model composed of Landau quasiparticle
states with patched Fermi surfaces (FS) sandwiched by states with flat
FS to simulate the ``cold'' spot regions in cuprates. We calculate the
one particle irreducible function and
the self-energy up to two-loop order. Using renormalization group arguments
we show that in the forward scattering channel
the renormalized coupling constant is never infrared stable due to the flat
FS sectors. Furthemore we show that the self-energy scales with energy as $%
\mathop{\rm Re}%
\sum \sim \omega {\it \ln }\omega $ as $\omega \rightarrow 0$, and thus
the Fermi liquid state within each FS patch is turned into a marginal Fermi
liquid.

\end{abstract}


The normal phase of the high-$T_c$ superconductors is by now notable for the
exhibition of a number of physical properties which don't fit in a
Landau-Fermi liquid framework. These anomalies have in common the fact that
they show a crossover behavior near optimal doping and they are in one
way or another deeply related with the peculiar nature of the electronic
excitations in the underdoped regime. In particular, the presence of an
anisotropic pseudogap in the underdoped regime is central to our
understanding of the FS topology in these materials. Based on the
photoemission spectroscopy data (ARPES) as well as other experimental
techniques \cite{Timusk} we have now a better view of the FS in the
cuprates. The experiments performed especially on underdoped Bi2212 and its
family compounds \cite{Shen} demonstrate the existence of a spectrum
characterized by the pseudo gap in the regions around $\left( \pm \pi
,0\right) $ and $\left( 0,\pm \pi \right) $ in the Brillouin zone. Besides
that the experimental data show a single-particle peak structure in the
ARPES spectra near $\left( \pm \frac \pi 2,\pm \frac \pi 2\right) $. It
emerges from these results the phenomenological picture of a two-dimensional
patched FS disconnected by pseudogap regions with the quasiparticle like
sectors locating in between, the so-called hot spots and cold spots
respectively \cite{Rice,sto}. The hot spots are in essence associated
with non-Fermi liquid 
like states presumably responsible for all the anomalies observed in the 
normal phase of the underdoped cuprates. In contrast, the cold spots are the
patches in which presumably Landau Fermi liquid theory holds.

Our purpose in this work is to test the validity of this scenario. In this
work we consider two-dimensional ($2d$) quasi-particles with a
truncated FS composed of four patches. The boundaries of each patch are
taken to be flat with a linear particle dispersion law. This latter
description is known to produce non-Fermi liquid behavior as the
result of FS nesting\cite{Yakovenko,vir}. The central regions of these patches
are occupied by conventional Fermi liquid states. We take into account
interactions between particles of opposite spins in both Cooper $\left(
C\right) $ and forward channels $\left( F\right) $. As expected we find
logarithmic divergent non-interacting susceptibilities in these two
channels. We calculate the resulting effective quasi-particle interaction up
to two-loop order. The Cooper channel is associated with the superconducting
instability and does not drive the physical system outside the Landau-Fermi
liquid regime for repulsive bare interactions. This is opposite to what
happens in the $F$ channel. These contributions are therefore momentum
dependent. Whenever the superconducting instability is dominant the
effective coupling constant approaches the Fermi liquid fixed point. If the
coupling constant is regulated by the $F$ channel instead it becomes
infrared divergent at the Fermi surface.

We also show that the flat sectors of the FS produce new features in the
self-energy $\Sigma $. We calculate the quasiparticle self-energy up to
two-loop order. The resulting $\sum $ scales as $\omega \ln \omega $ when $%
\omega \rightarrow 0$ , as a marginal Fermi liquid\cite{Varma}, producing $%
\frac{\partial {\it 
\mathop{\rm Re}%
}\sum }{\partial \omega }\rightarrow -\infty $ and a nullified quasiparticle
renormalization constant $Z=0$. As a result we conclude that there isn't a
truly stable Landau Fermi liquid in such a patched Fermi Surface if the
effects of its flat sectors are appropriately taken into account.

We consider initially a $2d$ Fermi Surface which consists of four
disconnected patches centered around $\left( \pm k_{F},0\right) $ and $%
\left( 0,\pm k_{F}\right) $ respectively, as shown in Fig. 1 (a). Let us
assume that they are all Landau Fermi liquid like. The disconnected arcs
separate occupied single-particle states from unoccupied ones. However as we
approach any patch boundary along the arc there cannot be such a sharp
separation of single-particle states. If we phrase this in terms of Fermi
liquid theory we can say that there are large incoherence effects near each
patch border which destroys the sharpness of the FS in these regions. That
is, the border regions within any patch cannot truly be Landau like. As a
result each one of the disconnected patches has al least finite segments
around its borders in which there should exist non-Fermi liquid states. We
choose to represent these non-Fermi liquid sectors by finite flat FS pieces
combining $2d$ and $1d$ features\cite{Anderson}. Our resulting FS model , as
shown in Fig. 1b, consists in this way of four symmetrical disconnected
patches as before with flat border sectors. The $2d$ Fermi liquid states
defined around the patch center are such that their dispersion law become
one-dimensional as we approach the borders since the FS curvature is
identically zero in these regions.

In order to make contact with conventional Fermi liquid theory let us
consider the Lagrangian density ${\cal L}$

\begin{equation}
{\cal L=}\sum_{\sigma }\psi _{\sigma }^{\dagger }\left( x\right) \left(
i\partial _{t}+\frac{\nabla ^{2}}{2m^{*}}+\mu \right) \psi _{\sigma }\left(
x\right) -\sum\limits_{\sigma ,\sigma ^{^{\prime }},\alpha ,\alpha
^{^{\prime }}}\int\limits_{y}U_{\sigma ,\sigma ^{^{\prime }};\alpha ,\alpha
^{^{\prime }}}\left( x-y\right) \psi _{\alpha }^{\dagger }\left( x\right)
\psi _{\alpha ^{^{\prime }}}^{\dagger }\left( y\right) \psi _{\sigma
^{^{\prime }}}\left( y\right) \psi _{\sigma }\left( x\right)
\end{equation}
where $\sigma ,\sigma ^{^{\prime }},\alpha ,\alpha ^{^{\prime }}$ are spin
indices and $\int\limits_{y}=\int dtd^{2}y$. Here $\mu =k_{F}^{2}/2m^{*}$
with the fermionic excitations represented by the fermion fields $\psi
_{\sigma }$ and $\psi _{\sigma }^{\dagger }$ well defined only in the region
defined by the four patches centred around the FS points $\left( \pm
k_{F},0\right) $ and $\left( 0,\pm k_{F}\right) $ ( see Fig. 1 (b) ).

Outside these disconnected patches the fermion fields are taken to be
identically zero. Following this argument the fermionic single-particle
energy $\varepsilon \left( {\bf p}\right) $ should be defined in accordance
with the patch under consideration. For the up and down patches defined
around the y-axis the Fermi liquid single-particle energy dispersion is
given by 
\begin{equation}
\varepsilon \left( {\bf p}\right) \cong \frac{k_{F}^{2}}{2m^{*}}+\left( \pm
p_{y}-k_{F}\right) v_{F}+\frac{p_{x}^{2}}{2m^{*}}
\end{equation}
with $k_{F}-\lambda \leq \pm p_{y}\leq k_{F}+\lambda $ and $-\Delta \leq
p_{x}\leq \Delta $, where $v_{F}$ is the Fermi velocity and $\lambda $ and $%
\Delta $ are wave vector cutoffs.

In the corresponding flat sectors the energy dispersion is $1d$. That is,

\begin{equation}
\varepsilon \left( {\bf p}\right) \cong \frac{k_{F}^{2}}{2m^{*}}+\left( \pm
p_{y}-k_{F}\right) v_{F}
\end{equation}

for $k_{F}-\lambda \leq \pm p_{y}\leq k_{F}+\lambda $ and $-\lambda \leq
p_{x}\leq -\Delta $ $or$ $\Delta \leq p_{x}\leq \lambda $.
We have similar results for the patches defined around $(\pm k_F,0)$
exchanging $p_x$ and $p_y$.

Within this framework we can use perturbation theory to calculate the
single-particle Green's function $G_{\sigma }$ , the particle-particle and
the particle-hole susceptibilities $\Pi _{\uparrow \downarrow }^{\left(
0\right) }$ and $\chi _{\uparrow \downarrow }^{\left( 0\right) }$
respectively and the one-particle irreducible functions $\Gamma _{\sigma
,-\sigma ;\sigma ,-\sigma }$ and $\Gamma _{\sigma ,\sigma ;\sigma ,\sigma }$%
. In all our perturbation theory calculations we assume for simplicity that
the interaction function $U\left( x\right) $ reduces to $U\delta \left(
x\right) $. 
%
%

Let us therefore proceed with our perturbation theory calculation of the
one-particle irreducible function $\Gamma _{\sigma ,-\sigma ;\sigma ,-\sigma
}\left( p_{1},p_{2};p_{3},p_{4}\right) $ up to two-loop order. The
corresponding diagrams for $\Gamma _{\uparrow \downarrow ;\downarrow
\uparrow }$ are drawn in Fig. 2. The diagram are given in terms of
particle-hole and particle-particle susceptibilities given by

%
\begin{eqnarray}
-i\chi _{\uparrow \downarrow }^{\left( 0\right) }\left( P\right)
&=&\int\limits_{q}G_{\uparrow }^{\left( 0\right) }\left( q\right)
G_{\downarrow }^{\left( 0\right) }\left( q+P\right) , \\
i\Pi _{\uparrow \downarrow }^{\left( 0\right) }\left( P\right)
&=&\int\limits_{q}G_{\uparrow }^{\left( 0\right) }\left( q\right)
G_{\downarrow }^{\left( 0\right) }\left( -q+P\right)
\end{eqnarray}
with the free single-particle Green's function $G^{\left( 0\right) }\left(
q\right) =$ $\frac{1}{q_{0}-(\varepsilon \left( {\bf q}\right) -\mu )\pm
i\delta }$ where $\varepsilon \left( {\bf q}\right) -\mu $ is calculated
around the central point of one of the four available FS patches being
either 1d or 2d depending on the relative position of the ${\bf q}$ vector.

Let us initially consider these scattering processes in the Cooper channel
which is characterized by the choice $p_{1}+p_{2}=\left( p_{0},0\right) $
with $p_{0}\rightarrow 0$. If we calculate the integrals for $i\Pi
_{\uparrow \downarrow }^{\left( 0\right) }\left( P\right) $ above we find in
this case

\begin{equation}
\Pi _{\uparrow \downarrow }^{\left( 0\right) }\left( p_{0}\right) \cong 
\frac{\lambda /\pi ^{2}}{k_{F}/m^{*}}\left[ \ln \left( \frac{2k_{F}\lambda
/m^{*}-p_{0}-i\delta }{p_{0}+i\delta }\right) +\ln \left( \frac{%
2k_{F}\lambda /m^{*}+p_{0}-i\delta }{p_{0}-i\delta }\right) \right]
\end{equation}
In this channel the ln singularity in $\Pi _{\uparrow \downarrow }^{\left(
0\right) }$ is produced by both flat and curved FL sectors.

Let us next calculate $\chi _{\uparrow \downarrow }^{\left( 0\right) }\left(
P\right) $ in the forward channel with ${\bf p}_{1}{\bf \ =p}_{3}{\bf ;p}_{2}%
{\bf =p}_{4}$ for $P=p_{4}-p_{1}=\left( p_{0},-2k_{F}{\bf y}\right) $. Using
the corresponding definition for $\chi ^{\left( 0\right) }$ we find that

\begin{equation}
\chi _{\uparrow \downarrow }^{\left( 0\right) }\left( P\right) \cong \frac{%
\left( \lambda -\Delta \right) m^{*}}{4\pi ^{2}k_{F}}[\ln \left( \frac{%
2k_{F}\lambda /m^{*}+p_{0}-i\delta }{p_{0}-i\delta }\right) +\ln \left( 
\frac{2k_{F}\lambda /m^{*}-p_{0}-i\delta }{p_{0}+i\delta }\right) ]
\end{equation}
As opposed to what happens with $\Pi _{\uparrow \downarrow }^{\left(
0\right) }\left( p_{0}\right) $ the singularity in $\chi ^{\left( 0\right) }$
is entirely produced by the flat sectors of the Fermi surface.

The appearance of singularities in $\Gamma $ is momenta dependent since it
is regulated by our choice of scattering channel. For ${\bf p}_{1}+{\bf p}%
_{2}={\bf 0}$ , $\Pi _{\uparrow \downarrow }^{\left( 0\right) }$ is singular
and $\chi _{\uparrow \downarrow }^{\left( 0\right) }$ is finite. For ${\bf p}%
_{4}-{\bf p}_{1}=-2k_{F}{\bf y}$ only $\chi _{\uparrow \downarrow }^{\left(
0\right) }$ is singular.

Thus in the Cooper channel for ${\bf p}_{1}+{\bf p}_{2}={\bf 0}$ it follows
that our perturbation theory expansion for $\Gamma _{\uparrow \downarrow
;\downarrow \uparrow }$ is dominated by the $\Pi _{\uparrow \downarrow
}^{\left( 0\right) }$ terms in the diagram series. We obtain in this case

\begin{equation}
\Gamma _{\uparrow \downarrow ;\downarrow \uparrow }\left( {\bf p}_{1}{\bf =-p%
}_{2}{\bf ;p}_{3}{\bf =-p}_{4};p_{0}\right) \cong -iU+iU\Pi _{\uparrow
\downarrow }^{\left( 0\right) }\left( p_{0}\right) +iU^{3}\left( ^{2}\Pi
_{\uparrow \downarrow }^{\left( 0\right) }\left( p_{0}\right) \right)
^{2}+...
\end{equation}

Following conventional renormalization theory procedure we can replace our
original coupling function by its bare analogue so that all divergences are
cancelled exactly in all orders of perturbation theory. For this let us
define the bare coupling function $U_{0\uparrow \downarrow ;\downarrow
\uparrow }\left( p_{1},p_{2};p_{3},p_{4}\right) $ such that if ${\bf p}_{1}+%
{\bf p}_{2}={\bf 0}$

\begin{equation}
U_{0\uparrow \downarrow ;\downarrow \uparrow }\left( {\bf p}_{1}{\bf \ =-p}%
_{2}{\bf ;p}_{3}{\bf =-p}_{4}\right) =U+U^{2}\frac{2\lambda /\pi ^{2}}{%
k_{F}/m^{*}}\ln \left( \frac{2k_{F}\lambda /m^{*}}{\omega }\right) +...
\end{equation}
with $\omega $ being an energy scale parameter which will eventually
approach zero. Using the condition $\omega \partial U_{0}/\partial \omega =0$
we obtain from this result the renormalization group$\left( \text{RG}\right) 
$ equation appropriate for a Fermi liquid\cite{Shankar} $
\beta \left( U\right) =\omega \frac{\partial U}{\partial \omega }=bU^{2} $%
where $b=\frac{2\lambda /\pi ^{2}}{k_{F}/m^{*}}$. This equation can be
easily integrated out to give

\begin{equation}
U\left( \omega \right) =\frac{U\left( \Omega \right) }{1+bU\left( \Omega
\right) \ln \left( \frac{\Omega }{\omega }\right) },
\end{equation}
where $\Omega $ is some fixed upper energy limit. Even if $U\left( \Omega
\right) $ is not small, $U\left( \omega \right) $ grows weaker for low
values of $\omega $ signalling the existence of a trivial Landau fixed point
for the renormalized $U_{\uparrow \downarrow ;\downarrow \uparrow }\left(
p_{1},p_{2};p_{3},p_{4}\right) $.

Consider next the case in which ${\bf P}={\bf p}_{4}-{\bf p}_{1}=-2k_{F}{\bf %
y}$ and ${\bf p}_{1}+{\bf p}_{2}=2\Delta {\bf x} $ with ${\bf p}_{1}{\bf \ =p%
}_{3}{\bf ;p}_{2}{\bf =p}_{4}$ and $p_{0}=\omega \rightarrow 0$. Using our
perturbation theory expansion we find

\begin{equation}
\Gamma _{\uparrow \downarrow ;\downarrow \uparrow }\left( {\bf p}_{1}{\bf =p}%
_{3}{\bf ;p}_{2}{\bf =p}_{4};\omega \right) \cong -iU-iU^{2}\chi _{\uparrow
\downarrow }^{\left( 0\right) }\left( \omega \right) +iU^{3}\left( \chi
_{\uparrow \downarrow }^{\left( 0\right) }\left( \omega \right) \right)
^{2}+...  \nonumber
\end{equation}
since for $2\Delta >\lambda $ , the $\chi _{\uparrow \downarrow }^{\left(
0\right) }\left( \omega \right) $ contributions dominate our perturbative
expansion above. Following the same route as before we can define the bare
interaction function $U_{0\uparrow \downarrow ;\downarrow \uparrow }\left(
p_{1},p_{2};p_{3},p_{4}\right) $ for ${\bf p}_{4}-{\bf p}_{1}$ $=$ ${\bf P}$
as

\begin{equation}
U_{0\uparrow \downarrow ;\downarrow \uparrow }\left( {\bf p}_{1}{\bf \ =p}%
_{3}{\bf ;p}_{2}{\bf =p}_{4};{\bf p}_{4}-{\bf p}_{1}={\bf P;}\omega \right)
=U-U^{2}\frac{m^{*}\left( \lambda -\Delta \right) }{2\pi ^{2}k_{F}}\ln
\left( \frac{2k_{F}\lambda /m^{*}}{\omega }\right) +...
\end{equation}
As a result in this channel if we neglect the tweo-loop order terms the
renormalized coupling $U$ satisfies the RG equation $
\beta \left( U\right) =\omega \frac{\partial U}{\partial \omega }=-cU^{2} $%
which when integrated out gives

\begin{equation}
U\left( \omega \right) =\frac{U\left( \Lambda \right) }{1-cU\left( \Lambda
\right) \ln \left( \frac{\Lambda }{\omega }\right) }
\end{equation}
for $c=\frac{m^{*}\left( \lambda -\Delta \right) }{2\pi ^{2}k_{F}}$ and $%
\Lambda $ being an upper energy limit. Clearly the physical system is driven
outside the domain of validity of perturbation theory if $cU\left( \Lambda
\right) \ln \left( \frac{\Lambda }{\omega }\right) \rightarrow 1$. The Fermi
liquid infrared fixed point is physically unattainable in this limit. In
fact if we consider the two-loop order terms the $\beta $ function series
becomes $
\beta \left( U\right) =-cU^{2}-aU^{3}+... $
where $a=2c^{2}\ln \left( \frac{2k_{F}\lambda /m^{*}}{\Omega }\right) $.
Thus there is a non-trivial fixed point $U^{*}=-1/2c\ln \left( \frac{%
2k_{F}\lambda /m^{*}}{\Omega }\right) .$ However since $\left( \partial
\beta \left( U\right) /\partial U\right) _{U^{*}}<0$ it is also infrared
unstable.

Consider next the calculation of the self-energy $\Sigma \left( p\right) $
up to two-loop order ( see Fig. 3 ). The one-loop diagram is trivial
producing only a constant shift in the Fermi energy. The same goes for the
two-loop diagram from Fig. 3 (b). However the remaining one gives a
nontrivial contribution: 
\begin{equation}
-i\Sigma _{\uparrow }\left( p\right) =-U^{2}\int\limits_{q}G_{\downarrow
}^{\left( 0\right) }\left( q\right) \chi_{\uparrow \downarrow }^{\left(
0\right) }\left( q-p\right)  \label{SE}
\end{equation}
From our previous calculation we noticed that the singularity in $\chi
^{\left( 0\right) }$ is produced by the flat sectors of the FS when there is
perfect nesting between the patches related by the scattering process. Since 
$p$ is an external variable we choose for convenience ${\bf p}=\left( \Delta
,-k_{F}\right) $. In this way the relevant patch for the ${\bf k}$
integration is $\left( 0,-k_{F}\right) $. Taking into account that ${\bf q}$
is located in the patch $\left( 0,k_{F}\right) $ we can evaluate $%
\chi_{\uparrow \downarrow }^{\left( 0\right) }\left( q-p\right)$ directly. 
If we replace the result in the Eq. (\ref{SE}) the self-energy reduces
simply to

\begin{equation}
\Sigma \left( p_{0}\right) =-\frac{U^{2}\left( m^{*}\left( \lambda -\Delta
\right) \right) ^{2}}{16\pi ^{4}\left( 2k_{F}\right) ^{2}}p_{0}\left[ \ln
\left( \frac{2k_{F}\lambda /m^{*}-p_{0}-i\delta }{p_{0}+i\delta }\right)
+\ln \left( \frac{2k_{F}\lambda /m^{*}+p_{0}-i\delta }{p_{0}-i\delta }%
\right) \right]
\end{equation}
Therefore it follows from this result that in the limit $p_{0}\rightarrow 0$
the self-energy $\Sigma $ reproduces the marginal Fermi liquid result 
${\rm 
\mathop{\rm Re}%
}\Sigma \left( p_{0}\right) \sim p_{0}\ln p_{0}$ 
producing in this way a charge renormalization parameter $Z$ identically
zero: $Z=\left( 1-\frac{\partial }{\partial p_{0}}{\rm 
\mathop{\rm Re}%
}\Sigma \left(p_{0}=0\right) \right) ^{-1}=0$. The Fermi liquid states are
in this way turned into a marginal Fermi liquid by the interaction effects
produced by the flat sectors of the patched Fermi surface.

In conclusion, in this work we test the validity of the phenomenological
picture of a patched FS composed of confined quasiparticle states (cold
spots) separated by non-Fermi liquid ( hot spots ) sectors which are
presumably responsible for the anomalous metallic properties observed in
the normal phase of underdoped high-temperature superconductors. We consider
a $2d$ disconnected Fermi surface composed of four different patches. In each
of these patches there are conventional Fermi liquid states defined around
their center as well as non-Fermi liquid states located in the border
sectors. These latter states are modeled by single-particle states with
flat Fermi surfaces. This representation is known to produce non-Fermi
liquid behavior due to the presence of perfect nesting of the Fermi
surface patches. For simplicity we assume that all particles interact
locally but we take proper account of the spin degrees of freedom. Using a
simplified model Lagrangian we calculate the one-particle irreducible
function $\Gamma_{\uparrow \downarrow ;\downarrow \uparrow }$up to two-loop
order. The perturbation series is momentum dependent. In the Cooper channel
the expansion is dominated by the particle-particle susceptibility $\Pi
_{\uparrow \downarrow }^{\left( 0\right) }$which is logarithmic divergent
even if flat sectors are not present in our system. Making a renormalization
group analysis we show that this channel is infrared stable with the
renormalized repulsive coupling constant approaching the Landau trivial
fixed point as we approach the Fermi surface. In contrast, in the
forward scattering channel the perturbation series is dominated by the
logarithmic divergence of $\chi _{\uparrow \downarrow }^{\left( 0\right) }$%
which is now produced by the flat sectors of FS. This time the first
derivative of the $\beta \left( U\right) $with respect to $U$is negative
definite. The system is now infrared unstable and only reaches a negative
non-trivial fixed point in the ultraviolet limit. Finally we calculate the
quasiparticle self-energy up two-loop order. We show that one of the second
order diagram produces a non-trivial contribution to $\Sigma $. When we
evaluate this term we find that it reproduces the marginal Fermi liquid
result ${\rm Re}\Sigma \sim \omega \ln \omega $when $\omega \rightarrow 0$.
Therefore we can conclude that the confined Fermi liquid states, the
so-called cold spots, are drastically altered and turned into a marginal
Fermi liquid by the quasiparticles located in the flat sectors of the Fermi
surface.

ACKNOWLEDGMENTS

Two of us ( AF and TS ) would like to acknowledge financial support from the
Conselho Nacional de Desenvolvimento Cientifico e Tecnologico (CNPq) and
Financiadora de Estudos e Projetos (FINEP).


\begin{figure}[h]
\caption{ Fermi surface patches: (a) Fermi liquid like states: (b) flat
sectors added to the curved patches; (c) local view of the $(0,{\bf k}_F)$
patch. }
\end{figure}

\begin{figure}[h]
\caption{$\Gamma_{\uparrow \downarrow ; \uparrow \downarrow}$ diagrams up to
two-loop order. }
\end{figure}

\begin{figure}[h]
\caption{ Self-energy $\Sigma_{\uparrow}$ up to two-loop order. }
\end{figure}


\begin{references}
\bibitem{Timusk}  {\it T.Timusk and B.Statt, preprint cond-mat/9905219}

\bibitem{Shen}  {\it Z.X.Shen and D.S.Dessau, Phys. Rep.253 (1995) 1}

\bibitem{Rice}  {\it R.Hlubina and T.Rice, Phys. Rev. B51( 1995 )9253.}
\bibitem{sto} {\it B.P.Stojkovic and D.Pines, Phys. Rev. Lett.76 (1996 )811; J.Schmalian, D.Pines and B.Stojkovic, Phys. Rev. Lett.80 ( 1998 ) 3839}

\bibitem{Yakovenko}  {\it A.T.Zheleznyak, V.M.Yakovenko and
I.E.Dzyaloshinskii, Phys. Rev. B55 ( 1998 ) 3200}.
\bibitem{vir} {\it A. Virosztek and
J.Ruvalds, Phys. Rev. B51(1990) 4064}.

\bibitem{Varma}  {\it C.M.Varma,P.B.Littlewood,S.Schmitt-Rink,E.Abrahams and
A. E. Ruckenstein, Phys. Rev. Lett.63(1989)1996}.

\bibitem{Anderson}  {\it P.W.Anderson, Phys. Rev. Lett. 65 (1990) 2306;
Phys.Rev.Lett64 (1990) 1839}.

\bibitem{Shankar}  {\it R.Shankar, Rev. Mod. Phys. 66 (1994) 129} 
\end{references}
\end{document}